\documentclass{article}[12pt]

\begin{document}
\setcounter{page}{1} \pagestyle{plain} \vspace{1cm}
\begin{center}
\Large{\bf Modified Gravity Model $f(Q,T)$ and Wormhole Solution}\\
\small \vspace{1cm} \vspace{0.5cm} {\bf S. Davood
Sadatian}$^{1,2}$, {\bf S. Mohamad Reza Hosseini}$^{1}$\\
\vspace{0.5cm} {\it $^{1}$Research Department of astronomy $\&$
cosmology, University of Neyshabur, P. O. Box 9319774446, Neyshabur,
Iran\\
$^{2}$ Department of Physics, Faculty of Basic Sciences, University
of Neyshabur, P. O. Box 9319774446, Neyshabur, Iran
 }\\email: {sd-sadatian@um.ac.ir   ,    sd-sadatian@neyshabur.ac.ir    ,    mrhosseiniy@yahoo.com}\\
\end{center}
\vspace{1.5cm}
\begin{abstract}
We investigate wormhole solutions using the modified gravity model
$f(Q,T)$ with viscosity and aim to find a solution for the existence
of wormholes mathematically without violating the energy conditions.
We show that there is no need to define a wormhole from exotic
matter and analyze the equations with numerical analysis to
establish weak energy conditions. In the numerical analysis, we
found that the appropriate values of the parameters can maintain the
weak energy conditions without the need for exotic matter. Adjusting
the parameters of the model can increase or decrease the rate of
positive energy density or radial and tangential pressures.
According to the numerical analysis conducted in this paper, the
weak energy conditions are established in the whole space if
$\alpha< 0$, $12.56 < \beta < 25.12$ or $\alpha > 0$, $\beta >
25.12$. The analysis also showed that the supporting matter of the
wormhole is near normal matter, indicating that the generalized
$f(Q,T)$ model with viscosity has an acceptable parameter space to
describe a wormhole without the need for exotic matter.
\\

\end{abstract}
\newpage

\section{Introduction}
A wormhole is a theoretical concept in physics that suggests the
existence of a shortcut or tunnel between two distant points in
space or time (Kanti et al, 2012; Lin et al., 2019; Parsaei et al.,
2020; Wang, 2022). It is often depicted as a tunnel that connects
two separate regions of the universe, allowing for faster travel or
even time travel. According to Einstein's theory of general
relativity, certain solutions of the equations allow for the
existence of wormholes. However, creating a traversable wormhole
would require exotic matter with negative energy density and large
negative pressure (Anchordoqui et al, 1997). This type of matter has
not been observed in nature so far. It is important to note that
wormholes are still purely theoretical and have not been proven to
exist. Scientists continue to study and explore the possibilities of
wormholes within the framework of theoretical physics (Kanti et al,
2012; Anchordoqui et al, 1997; Richarte et al, 2008; Lobo et al,
2009; Moraes et al, 2017a; Moraes et al, 2017b, Lin et al., 2019;
Parsaei et al., 2020; Wang, 2022). In 1935, Einstein and physicist
Nathan Rosen proposed the existence of "bridges" or "wormholes" that
could connect two separate points in space-time (Kanti et al, 2012).
These theoretical structures would require the presence of exotic
matter with negative energy density and negative pressure to keep
the wormhole stable and traversable. It is worth noting that
traversable wormholes, which would allow objects or information to
pass through them would require the presence of exotic matter with
properties that are not currently known to exist in nature (Lobo et
al, 2009). Additionally, wormholes are subject to various
theoretical challenges, such as the potential for instability, high
radiation, and the need for advanced technology to create and
sustain them (Lobo et al, 2009; Moraes et al, 2017a; Moraes et al,
2017b). Researchers continue to study the theoretical properties and
implications of wormholes within the framework of general relativity
and quantum physics, but more research is needed to determine if
they can exist in reality (Moraes et al, 2017a;
Moraes et al, 2017b; Parsaei et al., 2020; Wang, 2022).\\
Black Holes and Wormholes are distinct concepts in the field of
astrophysics and general relativity. While they share certain
similarities, such as their association with extreme gravity, they
have fundamental differences: A Black Hole is a cosmic object formed
from the collapse of massive stars or through other astrophysical
processes. It has an event horizon, a boundary beyond which nothing,
not even light, can escape the gravitational pull of the Black Hole.
The collapsed matter at the center of a Black Hole is known as a
singularity, where the laws of physics, as we currently understand
them, break down. Black Holes are characterized by their mass, spin
(angular momentum), and electric charge. They are observed as
regions in space with exceptionally strong gravitational effects,
often accompanied by the accretion of matter from surrounding
sources (Jamil et al, 2012; Gadbail et al., 2021). On the other
hand, a wormhole is a speculative concept in theoretical physics
that suggests the existence of a shortcut or passage between two
different regions of space-time. It is often depicted as a
tunnel-like structure connecting two distant points in the universe.
The stability of wormholes is a major challenge, as they are prone
to collapse or require precise arrangements of matter to maintain
their integrity. While Black Holes and Wormholes are distinct
concepts, there is a historical connection between them (Elizalde et
al, 2019). Further scientific research is needed to explore the
properties and potential existence
of wormholes in the universe.\\
$f(Q)$ gravity is a modified theory of gravity that extends the
symmetric teleparallel equivalent of General Relativity (GR) (Cai et
al, 2016). In $f(Q)$ gravity, the gravitational interaction is
driven by a nonmetricity scalar called $Q$. This theory has gained
attention in the field of cosmology due to its potential to explain
various phenomena, such as dark energy and the accelerated expansion
of the universe (Cai et al, 2016; Jimeneza et al, 2019; Jimenez et
al, 2018; Paliathanasis, 2023). Cosmological models based on $f(Q)$
gravity have shown efficiency in fitting observational data sets at
both the background and perturbation levels (Cai et al, 2016;
Jimeneza et al, 2019). Also, $f(Q)$ gravity provides a complete dark
energy scenario, where the nonmetricity scalar $Q$ describes the
gravitational interaction (Jimeneza et al, 2019; Jimenez et al,
2018; Solanki et al., 2021). We know that the $\Lambda CDM$ Universe
can be reconstructed in terms of e-folding in $f(Q)$ gravity
(Paliathanasis, 2023). Some models have been studied in $f(Q)$
gravity to understand its implications and compare them with
observational data (Xu et al, 2019; Paliathanasis, 2023; Shabani et
al, 2023; Calza et al, 2023; Albuquerque et al, 2022). Also, the
evolution of linear perturbations in $f(Q)$ gravity has been
investigated, with the design of $f(Q)$ to achieve specific
cosmological implications (Xu et al, 2019; Albuquerque et al, 2022;
Calza et al, 2023). Energy conditions have been explored in $f(Q)$
gravity, which falls under the class of teleparallel theories of
gravity (Cai et al, 2016). On the other hand, an extension of $f(Q)$
gravity is $f(Q,T)$ gravity (we use this model in the following),
where $T$ represents the trace of the energy-momentum tensor. This
extension introduces an extra force on massive particles due to the
coupling between $Q$ and $T$ (Xu et al, 2019). Following, the $f(Q,
T)$ modified gravity model with viscosity is chosen for
investigating wormholes due to its ability to provide alternative
explanations to standard gravity, and its cosmological consequences.
Therefore, it distinguishes it from other modified gravity models
that may have different modifications or additional parameters.
Also, bulk viscosity is the only viscous influence that can change
the background dynamics in a homogeneous and isotropic(anisotropic)
universe (confirmed by recent observational data). The role of
viscosity is in understanding the accelerated expansion of the
universe and its implications for wormholes. Besides, the $f(Q, T)$
modified gravity model with viscosity is used to investigate
the properties and behavior of wormholes, such as their stability, formation, and potential traversability. \\
Therefore, according to the above discussions, we organized the
structures of this work as follows: in Section 2, we consider a
modified $f(Q, T)$ gravity model with viscosity contain and
numerically investigates the wormhole solution in some special functions. Finally, in Section 3, the conclusion and summary are offered.\\
\section{Wormhole Solution in $f(Q, T)$ Gravity Model}
Einstein's theory of general relativity allows for the existence of
wormholes as a theoretical possibility, their actual existence in
the universe remains hypothetical. Modified gravity models, on the
other hand, introduce modifications to Einstein's theory of general
relativity to address certain unresolved issues or to incorporate
additional phenomena. These modifications can lead to different
predictions for the behavior and properties of wormholes compared to
those derived from general relativity. It is important to note that
the predictions of modified gravity models for wormholes can vary
depending on the specific modifications introduced and the
assumptions made in those models. Some modified gravity models may
propose alternative explanations for wormholes or suggest different
properties and behaviors for these hypothetical structures.
Altogether, both Einstein's theory of general relativity and
modified gravity models allow for the possibility of wormholes, but
their predictions and implications can differ. There are both
discrepancies and points of convergence when comparing the
predictions for wormholes in modified gravity models with those
derived from Einstein's theory of general relativity or other
established models. 1)Discrepancies: Modified gravity models
introduce modifications to general relativity, which can lead to
different predictions for the behavior and properties of wormholes
compared to those derived from general relativity. Wormholes in
modified gravity models may not require exotic matter to prevent
collapse, unlike in general relativity. 2)Points of Convergence:
Both general relativity and modified gravity models allow for the
possibility of wormholes as theoretical constructs. Wormholes are
solutions to the Einstein field equations for gravity, which are the
foundation of general relativity (Lin et al.,
2019; Parsaei et al., 2020; Wang, 2022).\\
One of the fascinating aspects of $f(Q,T)$ gravity is its potential
to provide solutions for the existence of wormholes without the need
for exotic matter. In the following, the goal is to find solutions
that do not violate energy conditions and are consistent with the
laws of physics.\\
\subsection{Generalized Gravity Model $f(Q,T)$}
To better understand the nature of wormholes and their theoretical
underpinnings, we have explored modified theories of gravity. One
such theory is $f(Q,T)$ gravity, an extension of the symmetric
teleparallel equivalent of General Relativity (Cai et al, 2016). In
$f(Q,T)$ gravity, the gravitational interaction is driven by a
nonmetricity scalar called $Q$, which offers potential explanations
for phenomena like dark energy and the accelerated expansion of the
universe [Cai et al, 2016; Jimeneza et al, 2019; Xu et al, 2019). By
generalizing the gravity model $f(Q)$, we can get the gravity model
$f(Q,T)$, the action of this model is expressed as follows (Jimeneza
et al, 2019)
\begin{equation}
S=\int d^{4} x(\frac{1}{16\pi } f(Q,T)+L_{m} )\sqrt{-g}
\end{equation}
where $Q$ is the nonmetricity parameter, $T$ is the trace
energy-momentum tensor, $L_m$ stands for the matter Lagrangian and
$g=det(g_{\mu\nu})$. If we take the derivative of this action with
respect to the metric and connection, the equations of motion are
obtained as follows
$$-\frac{2}{\sqrt{-g} } \nabla _{\lambda } (\sqrt{-g} f_{Q} P^{\lambda } {}_{\mu \nu } )-\frac{1}{2} fg_{\mu \nu } +f_{T} (T_{\mu \nu } +\theta _{\mu \nu } )-f_{Q} (P_{\nu \rho \sigma } Q_{\mu } {}^{\rho \sigma } -2P_{\rho \sigma \mu } Q^{\rho \sigma } {}_{\nu } )=8\pi T_{\mu \nu
}$$
\begin{equation}
\nabla _{\mu } \nabla _{\nu } (\sqrt{-g} f_{Q} P_{\alpha }^{\mu \nu
} +4\pi H_{\alpha }^{\mu \nu } )=0
\end{equation}
where $f_{T} =\frac{\partial f}{\partial T} $, non-metricity tensor
$Q_{\lambda\mu\nu}=\nabla_\lambda g_{\mu\nu}$, $P^\lambda_{\mu\nu}$
is the superpotential, $f_{Q} =\frac{\partial f}{\partial Q} $, and
$H_{\alpha }^{\mu \nu } $ is the density of the hypermomentum and
are defined as follows (Jimeneza et al, 2019)
\begin{equation}
H_{\lambda }^{\mu \nu } \equiv \frac{\sqrt{-g} }{16\pi } f_{T}
\frac{\delta T}{\delta \hat{\Gamma }_{\mu \nu }^{\lambda } }
+\frac{\delta \sqrt{-g} L_{M} }{\delta \hat{\Gamma }_{\mu \nu
}^{\lambda } }
\end{equation}
where $\hat{\Gamma }_{\mu \nu }^{\lambda}$ is the Levi-Civita
connection. To calculate $f_{T} (T_{\mu \nu } +\theta _{\mu \nu } )$
that appeared in the equations of motion, we need to specify the
tensors $T_{\mu \nu } ,\theta _{\mu \nu } $ in the model. The
momentum energy tensor and $\theta _{\mu \nu } $ tensor are defined
as follows
\begin{equation}
T_{\mu \nu } \equiv -\frac{2}{\sqrt{-g} } \frac{\delta (\sqrt{-g}
L_{M} )}{\delta g^{\mu \nu }}
\end{equation}
\begin{equation}
\theta _{\mu \nu } =g^{\alpha \beta } \frac{\delta T_{\alpha \beta }
}{\delta g^{\mu \nu }}.
\end{equation}
\subsection{Wormhole Solution in $f(Q,T)$  Model with Viscosity}
It is worth mentioning that wormhole physics is a subject of ongoing
research and exploration within the field of theoretical physics.
Wormholes are solutions derived from the field equations in
Einstein's theory of gravitation, and their mathematical description
involves concepts from general relativity and differential geometry.
However, wormhole solutions have been studied using various
mathematical approaches and techniques. Some studies have explored
wormholes with variable equations of state (EoS) parameters, where
the violation of energy conditions is minimized or controlled (that
we use in our paper). We explored wormholes with variable equations
of state (EoS) parameters, where the ratio of pressure to energy
density $(\omega=\frac{p}{\rho})$ is a function of the
radial/tangential coordinate and we found a specific situation where
the energy conditions are satisfied. Other researchers have used the
cut-and-paste method to find wormhole solutions of finite size that
minimize the violation of energy conditions (Solanki et al., 2021).
In summary, wormholes are solutions derived from the field equations
in Einstein's theory of gravitation, and their mathematical
description involves concepts
from general relativity and differential geometry.\\
To calculate the solution of the wormhole, we assume the metric to
be static and have spherical symmetry. Using the Schwarzschild
coordinates $(t,r,\varphi ,\theta )$, we have
\begin{equation}
ds^{2} =(e^{2\phi (r)} )dt^{2} -(1-\frac{b(r)}{r} )^{-1} dr^{2}
-(r^{2} )d\theta ^{2} -(r^{2} \sin ^{2} \theta )d\varphi ^{2}
\end{equation}
where $b(r), \phi (r)$ are shape function and redshift function
respectively. We know that wormhole solution in any gravity model
should apply in the following conditions (Jamil et al, 2012; Kanti
et al, 2012; Anchordoqui et al, 1997; Richarte et al, 2008; Lobo et
al,
2009; Moraes et al, 2017a; Moraes et al, 2017b):\\
1) In the state $r>r_{0}$, shape function should be to form
$b(r)<r$, But in the throat of
the wormhole $r=r_{0} $, it must be $b(r_{0} )=r_{0} $.\\
2) The shape function must satisfy the flaring out, this means
$b'(r_{0} )<1$.\\
3) We need the condition of being asymptotically flat when $r\to
\infty $ then $\frac{b(r)}{r} \to 1$\\
4) The redshift function $\phi (r)$ must be bounded at all points in space.\\
Due to the pressure difference in the tangential and radial
direction, the condition of anisotropy is established, so the
momentum energy tensor is expressed as follows
\begin{equation}
T_{\mu }^{\nu } =(\rho +P_{t} )U_{\mu } U^{\nu } -P_{t} \delta _{\mu
}^{\nu } -(P_{r} -P_{t} )V_{\mu } V^{\nu }
\end{equation}
where $\rho, P_{r} ,P_{t} $ are the density of the universe, the
radial pressure and the tangential pressure and are a function of
radial coordinate $r$. $U_\mu$ and $V_\mu$ are the four-velocity
vector and unitary space-like vectors. Also, the trace of the
momentum energy tensor will be $T=\rho -P_{r} -2P_{t} $. If we
consider the Lagrangian of the matter according to (Richarte et al,
2008; Lobo et al, 2009; Moraes et al, 2017a; Moraes et al, 2017b;
Jamil et al, 2012) to be equal to $L_{m} =P$, the equation (5)
becomes as follows
\begin{equation}
\theta _{\mu \nu } =-g_{\mu \nu } P-2T_{\mu \nu }
\end{equation}
where $P=\frac{P_{r} +2P_{t} }{3} $. The nonmetricity according to
the spherically symmetric metric is as follows
\begin{equation}
Q=\frac{-b}{r^{2} } [\frac{rb'-b}{r(r-b)} +2\phi']
\end{equation}
Now we can calculate the equations of motion by using the metric and
tensor of momentum energy and the tensor of nonmetricity. Therfore,
by using equations (9), (7) and (6) ,and inserting them in the
equation of motion (2), we have

$$8\pi \rho =\left(\frac{r-b}{2r^{3} } \right)[f_{Q}
\left(\frac{\left(2r-b\right)\left(rb'-b\right)}{\left(r-b\right)^{2}
} +\frac{\left(2+2r\phi '\right)b}{r-b} \right)$$
\begin{equation}
+f_{QQ} Q'\left(2\frac{br}{r-b} \right)+f\left(\frac{r^{3} }{r-b}
\right)+f_{T} \left(\frac{2r^{3} }{r-b} \right)\left(P+\rho \right)]
\end{equation}

$$8\pi P_{r} =-\left(\frac{r-b}{2r^{3} } \right)[f_{Q}
\left(\frac{b}{r-b} \left(\frac{\left(rb'-b\right)}{r-b} +2+2r\phi
'\right)-4r\phi '\right)$$
\begin{equation}
+\frac{2br}{r-b} f_{QQ} Q'+f\left(\frac{r^{3} }{r-b}
\right)-2\left(\frac{r^{3} }{r-b} \right)f_{T} \left(P-P_{r}
\right)]
\end{equation}

$$8\pi P_{t} =-\left(\frac{r-b}{4r^{2} } \right)[f_{Q}
\left(\frac{\left(rb'-b\right)\left(\frac{2r}{r-b} +2r\phi
'\right)}{r\left(r-b\right)} +\frac{4\left(2b-r\right)\left(\phi
'\right)}{\left(r-b\right)} -4r\left(\phi '\right)^{2} -4r\phi
''\right)$$
\begin{equation}
-4rf_{QQ} Q'\phi '+\frac{2fr^{2} }{r-b} -\frac{4r^{2} f_{T}
\left(P-P_{t} \right)}{r-b} ]
\end{equation}
where prime denotes derivative to $r$. To solve the equations of
motion, the function $f(Q,T)$ of the model must be known. According
to (Xu et al, 2019), we consider the gravitational function $f(Q,T)$
of our model as $f(Q,T)=\alpha Q^{-1} +\beta T$ ($\alpha$ and
$\beta$ are arbitrary parameters). Also, we choose shape and redsift
functions as (Tayde et al., 2023)
\begin{equation}
b(r)=r_{0} (\frac{r_{0} }{r} )^{n}
\end{equation}
\begin{equation}
\phi (r)=\phi _{0} (\frac{r_{0} }{r} )^{m}
\end{equation}
where $r_{0} $ is a positive constant and  $\phi _{0} $ is an
arbitrary constant value and $n$,$m$ are constant exponents which
are strictly positive to satisfy the asymptotic flatness condition
(Tayde et al., 2023). Therefore, with these functions, we will have
three equations of motion and three unknown cosmic parameters
that we can solve.\\
But one of our assumptions in this article is the presence of
viscosity in the universe, so the pressure in the radial and
tangential direction is obtained according to the following
equations (Wang, 2022; Parsaei et al, 2022)
\begin{equation}
P_{r} {}_{v} =P_{r} -3\xi H_{0}
\end{equation}
\begin{equation}
P_{t} {}_{v} =P_{t} -3\xi H_{0}
\end{equation}
which $\xi $ (viscosity parameter) is defined as follows
\begin{equation}
\xi =\xi _{0} +\xi _{1} H_{0}
\end{equation}
where $H_{0} =73.24 \,kms^{-1}Mpc^{-1}$ and $\xi_{0}\simeq 10^{-6}$
(Wang, 2022), for simplicity we consider $\xi _{1} =0$ (Here we used
a similar procedure in (Wang, 2022) to introduce the viscosity
effect $-3\xi H_{0}$ in the equations). Now by using above equations
in and the equations (11) and (12), the field equations for
tangential and radial pressure in the case that the wormhole has
viscosity will be as follows
$$
8\pi P_{rv} =-\left(\frac{r-b}{2r^{3} } \right)[f_{Q}
\left(\frac{b}{r-b} \left(\frac{\left(rb'-b\right)}{r-b} +2+2r\phi
'\right)-4r\phi '\right)
$$
\begin{equation}
+\frac{2br}{r-b} f_{QQ} Q'+f\left(\frac{r^{3} }{r-b}
\right)-2\left(\frac{r^{3} }{r-b} \right)f_{T} \left(P-P_{r}
\right)]
\end{equation}
$$
8\pi P_{tv} =-\left(\frac{r-b}{4r^{2} } \right)[f_{Q}
\left(\frac{\left(rb'-b\right)\left(\frac{2r}{r-b} +2r\phi
'\right)}{r\left(r-b\right)} +\frac{4\left(2b-r\right)\left(\phi
'\right)}{\left(r-b\right)} -4r\left(\phi '\right)^{2} -4r\phi
''\right)
$$
\begin{equation}
-4rf_{QQ} Q'\phi '+\frac{2fr^{2} }{r-b} -\frac{4r^{2} f_{T}
\left(P-P_{t} \right)}{r-b} ]
\end{equation}
For the simplicity of calculations, we simplify equations and ignore
powers higher than $r^{3} $. Therefore, we have
\begin{equation}
\rho =\frac{r^3\alpha \beta {{\phi}_0}^2}{24{r_0}(32{\pi }^2-12\pi
\beta +{\beta }^2)}
\end{equation}
\begin{equation}
P_{rv}=-\frac{5\pi r^3\alpha \beta {{\phi}_0}^2}{3{r_0}(4\pi -\beta
)(8\pi -\beta )(8\pi +\beta )}+\frac{7r^3\alpha {\beta
}^2{{\phi}_0}^2}{24{r_0}(4\pi -\beta )(8\pi -\beta )(8\pi +\beta )}
\end{equation}
\begin{equation}
P_{tv}=-\frac{8{\pi }^2r^3\alpha {{\phi}_0}^2}{{r_0}(8\pi +\beta
)(32{\pi }^2-12\pi \beta +{\beta }^2)}+\frac{4\pi r^3\alpha \beta
{{\phi}_0}^2}{3{r_0}(8\pi +\beta )(32{\pi }^2-12\pi \beta +{\beta
}^2)}+\frac{r^3\alpha {\beta }^2{{\phi}_0}^2}{24{r_0}(8\pi +\beta
)(32{\pi }^2-12\pi \beta +{\beta }^2)}
\end{equation}
Now, we can check the condition of weak energy for the existence of
wormhole solution despite the presence of viscosity. According to
the condition of weak energy, the following three conditions should
be satisfied
\begin{equation}
\rho \ge 0\, \, \, \, , \, \, \, \rho +P_{rv} \ge 0\, \, \,, \,
\,\,\rho +P_{tv} \ge 0\,
\end{equation}
Also, we should check the border equation $r_{0} $ in the wormhole
throat. According to the boundary condition $r=r_{0} $, we have
\begin{equation}
\rho +P_{tv}=-\frac{{{r_0}}^2\alpha (24\pi +\beta
){{\phi}_0}^2}{768{\pi }^2-12{\beta }^2}
\end{equation}
\begin{equation}
\rho +P_{rv}=-\frac{{{r_0}}^2\alpha \beta {{\phi}_0}^2}{192{\pi
}^2-3{\beta }^2}
\end{equation}
For checking the condition of weak energy (23), we have the
following equations
\begin{equation}
\rho =\frac{r^3\alpha \beta {{\phi}_0}^2}{24{r_0}(32{\pi }^2-12\pi
\beta +{\beta }^2)}\ge 0
\end{equation}
\begin{equation}
\rho +P_{tv}=-\frac{r^3\alpha (24\pi +\beta
){{\phi}_0}^2}{12{r_0}(8\pi -\beta )(8\pi +\beta )}\ge 0
\end{equation}
\begin{equation}
\rho +P_{rv}=-\frac{r^3\alpha \beta {{\phi}_0}^2}{192{\pi
}^2{r_0}-3{r_0}{\beta }^2}\ge 0
\end{equation}
By determining the sign of equations (26,27,28), we obtain the
allowed intervals for the values $\alpha$ and $\beta$ so that the
boundary condition and the weak energy condition are established. In
this regard, we have
\begin{equation}
\alpha <0\, \, \, ,\, \, \, 4\pi <\beta <8\pi
\end{equation}
\begin{equation}
\alpha >0\, \, \, ,\, \, \, \beta >8\pi \,
\end{equation}
Using equations (20,21,22) and according to the equation of state
$\rho =\frac{P}{\omega } =(\frac{P_{rv} +2P_{tv} }{3\omega } )$, we
have
\begin{equation}
\rho =-\frac{r^3\alpha (16\pi -3\beta
){\phi_0}^2}{24r_0\omega(32{\pi }^2-12\pi \beta +{\beta }^2)}
\end{equation}
\begin{equation}
\rho +P_{tv}=-\frac{r^3\alpha (64{\pi
}^2(2+3\omega)-(3+\omega){\beta }^2-8\pi (\beta +4\omega\beta
)){\phi_0}^2}{24r_0}\omega(8\pi +\beta )(32{\pi }^2-12\pi \beta
+{\beta }^2)
\end{equation}
\begin{equation}
\rho +P_{rv}=-\frac{r^3\alpha (128{\pi }^2+8\pi (-1+5\omega)\beta
-(3+7\omega){\beta }^2){\phi_0}^2}{24r_0}\omega(4\pi -\beta )(8\pi
-\beta )(8\pi +\beta )
\end{equation}
Now, using these equations, we can check the different behaviors of
the wormhole according to the different $\omega$ values. To further
understand the conditions for establishing traversable wormholes, we
have employed numerical analysis techniques. By analyzing the
equations in $f(Q,T)$ gravity model, we determine and study the
implications of weak energy conditions. These investigations offer
valuable insights into the potential existence and properties of
wormholes within the framework of $f(Q,T)$
gravity.\\
\subsection{Numerical Analysis of Equations }
Energy conditions are important in studying the properties of
spacetime and the matter sources that generate it. For example, the
violation of the NEC in a wormhole solution would indicate the
presence of exotic matter in the throat of the wormhole.
Additionally, a positive energy density is required for a physically
realistic matter source that can sustain a wormhole solution in
general relativity.
\begin{figure}
\begin{center}\includegraphics{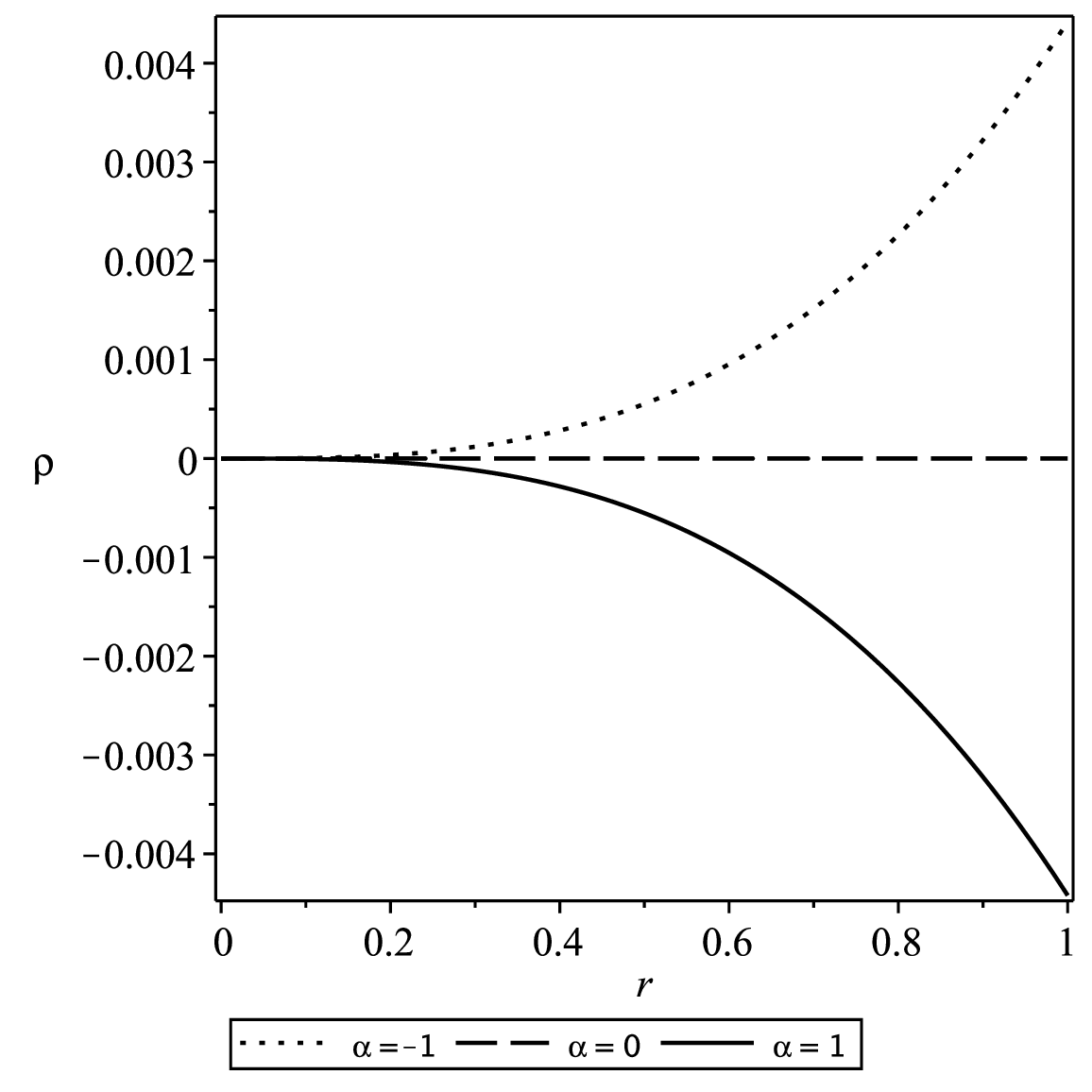} \vspace{5cm}\
\end{center}
\caption{\small {Plot of parameter $\rho$ (eV) versus $r$ (m) for
$r_0=2 , \phi_0=-1$ , $ \beta=15, \omega=-1$ and $\alpha=-1,0,1$ (we
use units $\hbar=c=1$ in our calculations and some parameters are
dimensionless).}}
\end{figure}

\begin{figure}
\begin{center}\includegraphics{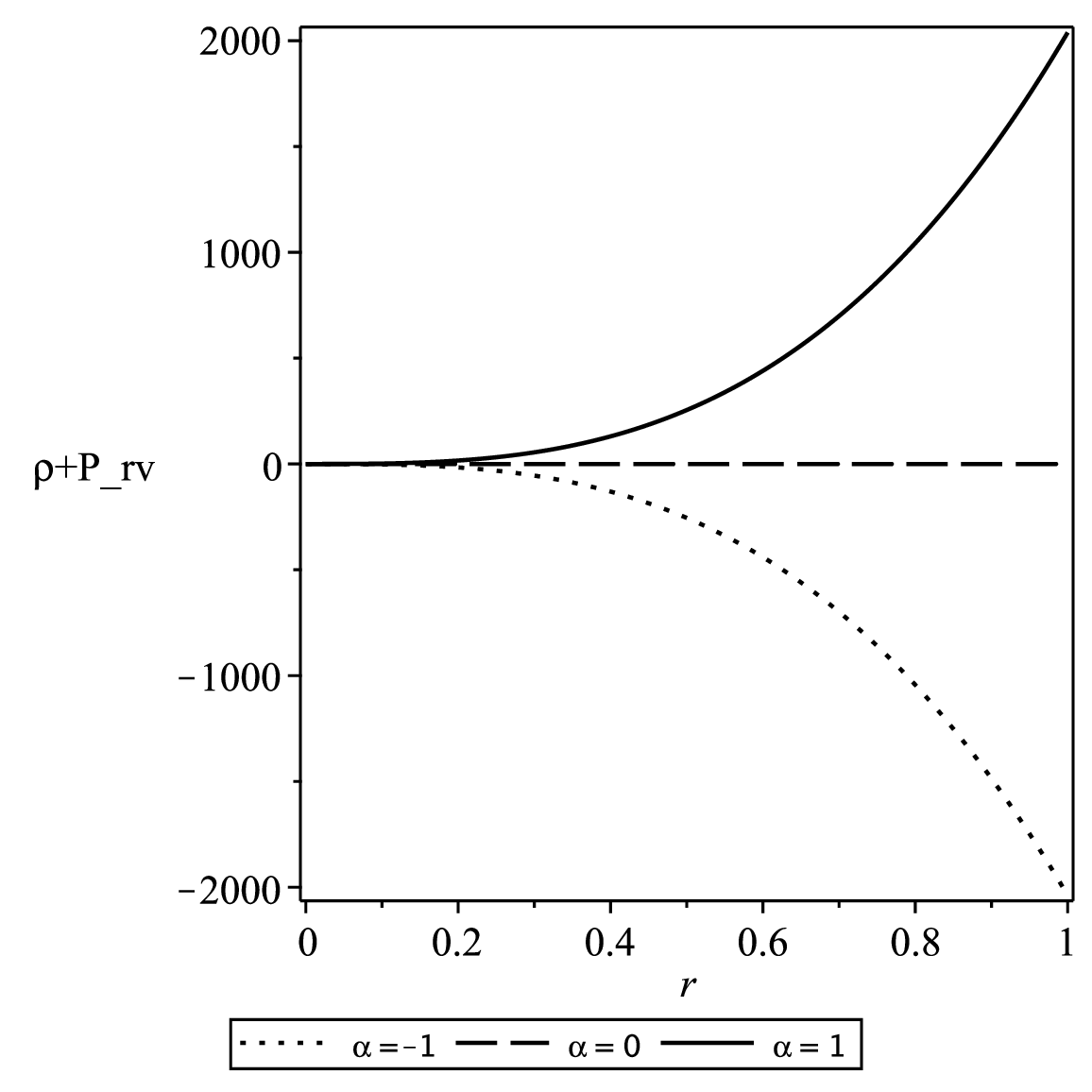} \vspace{4.5cm}\
\end{center}
\caption{\small {Plot of parameter ($\rho+P_{rv}$) versus $r$ (m)
for $r_0=2 , \phi_0=-1$ , $ \beta=15, \omega=-1$ and
$\alpha=-1,0,1$.}}
\end{figure}

\begin{figure}
\begin{center}\includegraphics{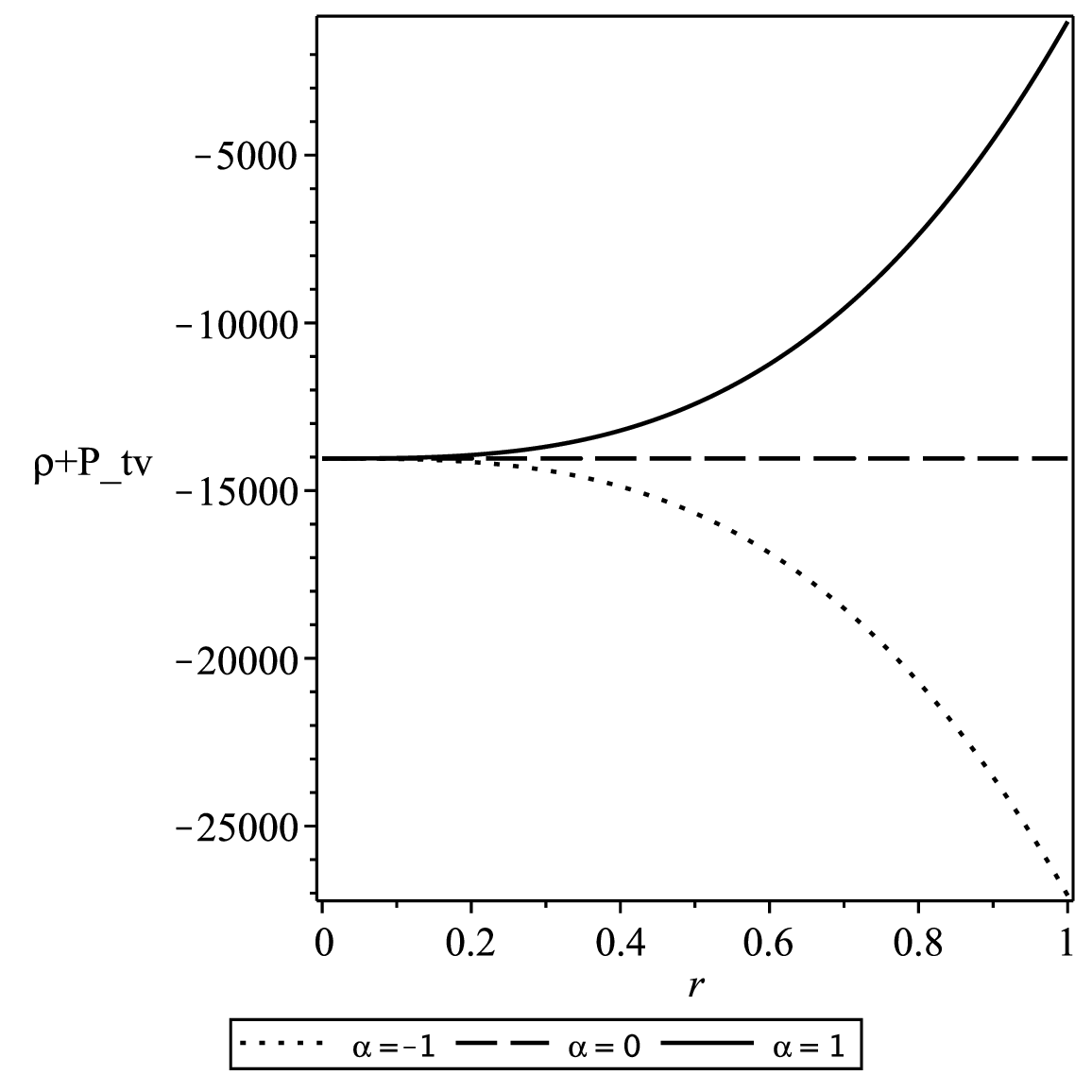} \vspace{5.8cm}\
\end{center}
\caption{\small {Plot of parameter ($\rho+P_{tv}$) versus $r$ (m)
for $r_0=2 , \phi_0=-1$ , $ \beta=15, \omega=-1$ and
$\alpha=-1,0,1$.}}
\end{figure}

\begin{figure}
\begin{center}\includegraphics{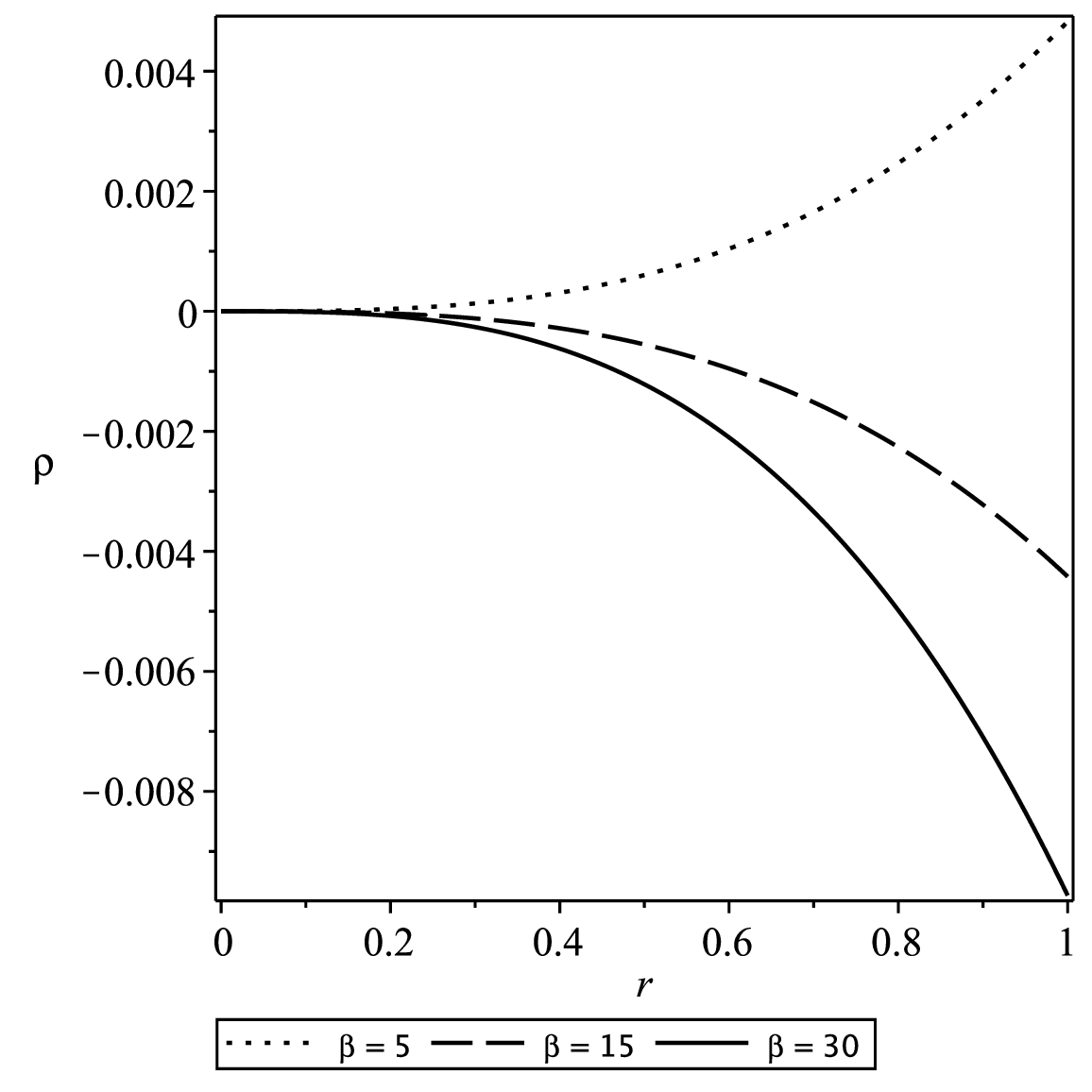} \vspace{5.8cm}\
\end{center}
\caption{\small {Plot of parameter ($\rho$) versus $r$ (m) for
$r_0=2 , \phi_0=-1$ , $ \alpha=1, \omega=-1$ and $\beta=5,15,30$.}}
\end{figure}

\begin{figure}
\begin{center}\includegraphics{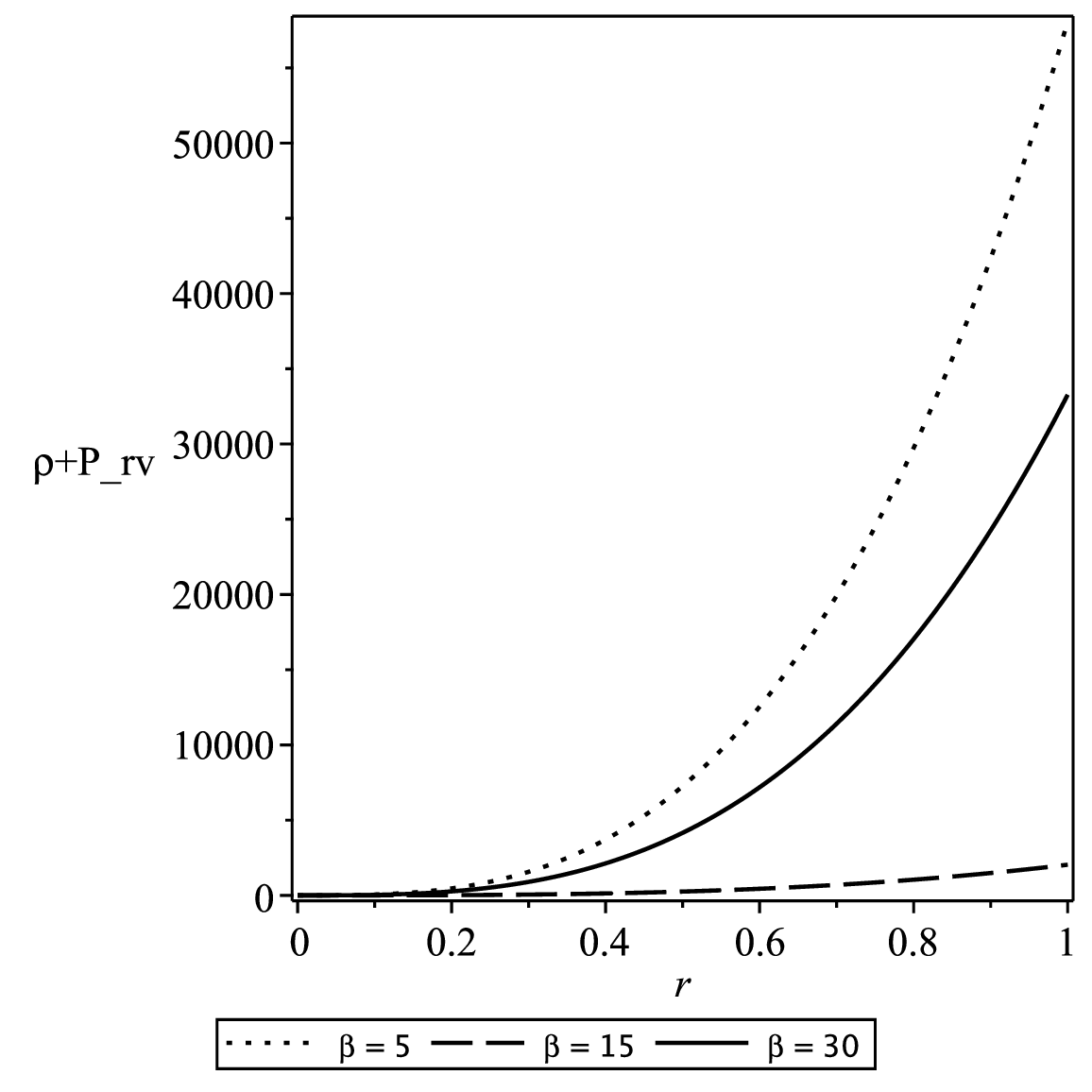} \vspace{4cm}\
\end{center}
\caption{\small {Plot of parameter ($\rho+P_{rv}$) versus $r$ (m)
for $r_0=2 , \phi_0=-1$ , $ \alpha=1, \omega=-1$ and
$\beta=5,15,30$.}}
\end{figure}

\begin{figure}
\begin{center}\includegraphics{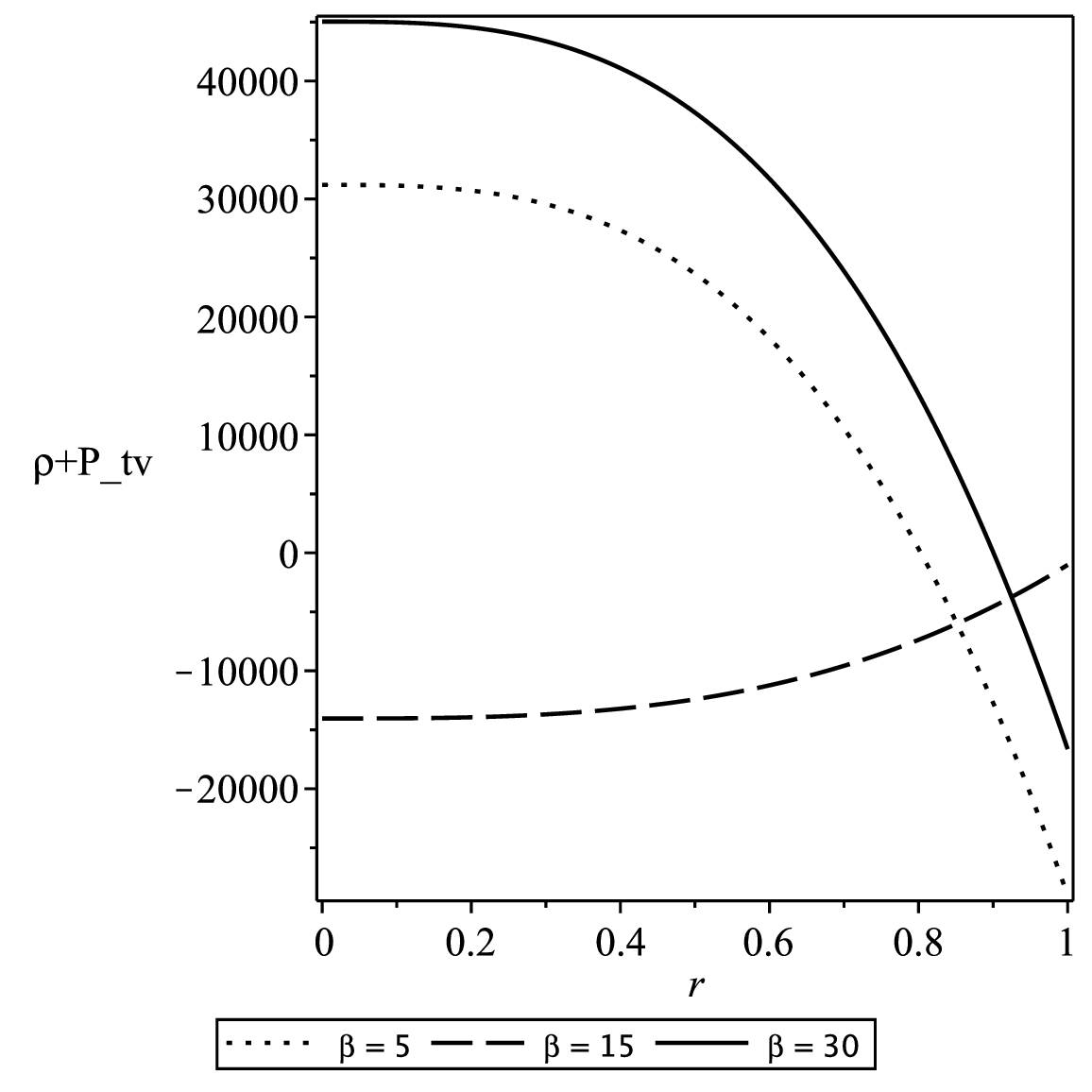} \vspace{6cm}\
\end{center}
\caption{\small {Plot of parameter ($\rho+P_{tv}$) versus $r$ (m)
for $r_0=2 , \phi_0=-1$ , $ \alpha=1, \omega=-1$ and
$\beta=5,15,30$.}}
\end{figure}

\begin{figure}[htp]
\begin{center}\includegraphics{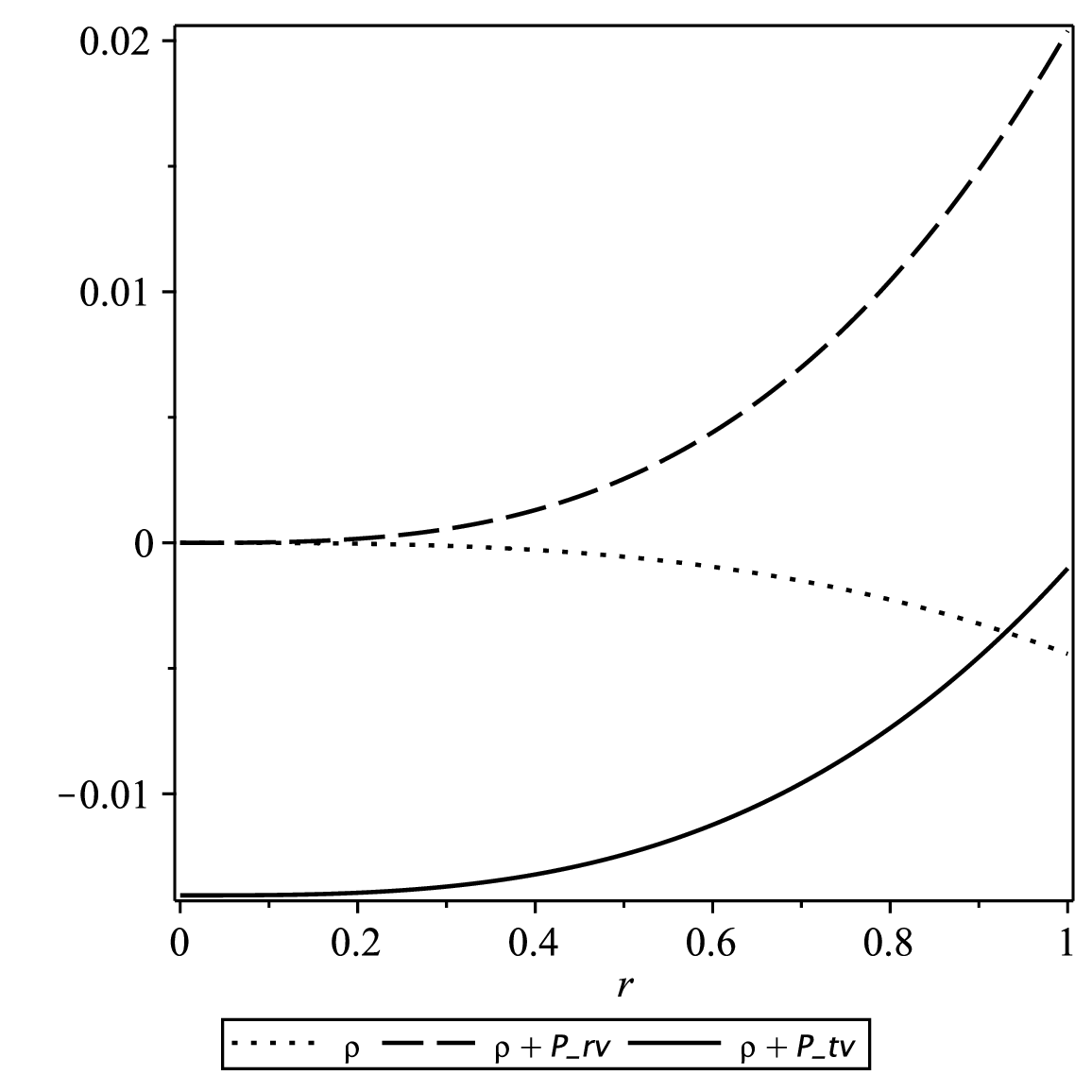} \vspace{6.5cm}\includegraphics{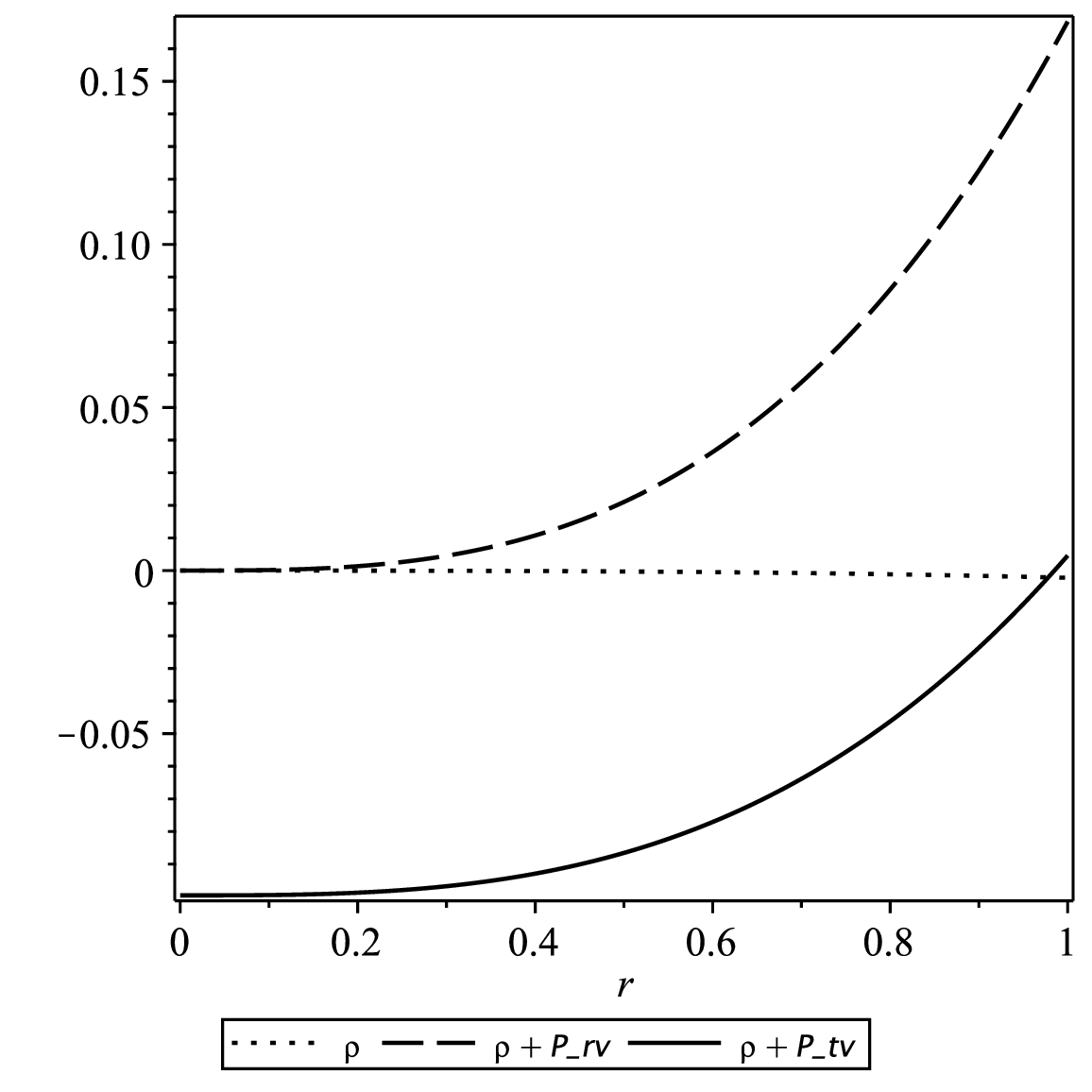}
\end{center}
\caption{\small {Variation of  parameters ($\rho, P_{rv},P_{tv}$)
relative to $r$ (m) with $\omega=-1$ (left),  $\omega=-2$ (right)
for $r_0=2, \phi_0=-1, \alpha=1,  \beta=15$. *we have rescaled the
graphs for better interpretation* }}
\end{figure}
In the following, we check the weak energy conditions using the
field equations. Based on this, in figures (1-7), the graphs of
$\rho+P_{rv}, \rho+P_{tv}, \rho$ have been drawn in terms of $r$,
where we have assumed $r_0=2 , \phi_0=-1$. We know that due to the
positiveness of $\rho$ and $\rho+P_r$, weak energy conditions are
established in the whole space, and if $\rho$ is negative, the weak
energy conditions are violated in the whole space. Therefore,
according figures (1-7), we showed that based on the appropriate
values of the parameters, weak energy condition is maintained if
$\alpha <0 ~,~ 12.56 <\beta <25.12$ or $\alpha
>0 ~,~ \beta >25.12 $. On the other hand, the
adjustment of some parameters is necessary to increase or decrease
the rate of positive energy density or radial and tangential
pressures. Also, according to the figure (7), it is clear that if
$\omega$ becomes more negative and $\alpha$ becomes more positive,
the energy density increases at a slower rate, and the energy
density becomes more positive. Therefore, the supporting matter of
the wormhole is near normal matter. \\
Finally, it can be concluded that the generalized $f(Q,T)$ model
with viscosity  has acceptable parameters space to describe a
wormhole without the need for exotic matter. That is, by accurately
setting the parameters in this model, the formation and description
mechanism of a wormhole can be explained. The proposed $f(Q, T)$
modified gravity model with viscosity offers a potential solution to
eliminate the need for exotic matter in wormhole formation. Exotic
matter, such as matter with negative mass, has been theorized to be
necessary to stabilize wormholes. However, the $f(Q, T)$ model
suggests an alternative approach. In the present $f(Q, T)$ modified
gravity model, the inclusion of viscosity, plays a crucial role in
understanding the behavior of wormholes. By introducing viscosity,
the model explores the effects of the accelerated expansion of the
universe on wormholes. Also, This implies that the stability and
formation of wormholes can be explained within the framework of the
$f(Q, T)$ model without the need for exotic matter. This means, the
influence of viscosity, the model explores the behavior and
characteristics of wormholes within the framework of modified
gravity. Here, we only studied the weak energy condition (WEC) for
model $f(Q,T)$, but the condition Null energy (NEC), Strong energy
condition (SEC), and Dominant energy condition
(DEC) can also be checked with the above method.\\
However, understanding wormholes without the need for exotic matter
could have several potential practical implications or applications
such as 1)Efficient Space Travel, 2)Time Travel Possibilities,
3)Cosmological Insights(Studying wormholes and their properties
could provide valuable insights into the fundamental nature of
spacetime and the laws of physics. It could help refine our
understanding of gravity, quantum mechanics, and the nature of the
universe), 4)Advanced Communication, and finally Exploring
Fundamental Physics(Investigating wormholes without the need for
exotic matter could shed light on the nature of matter, energy, and
the fundamental forces of the universe. It could contribute to the
development of new theories and models that go beyond our current
understanding of physics).\\
\section{Conclusion and Summary}
Wormholes offer the possibility of shortcuts or tunnels between two
distant points in space or time. While these theoretical constructs
are a product of Einstein's theory of general relativity, the
existence of traversable wormholes would require the presence of
exotic matter with negative energy density and negative pressure. In
this paper, we presented a modified gravity model $f(Q,T)$ with
viscosity that offers a potential solution to eliminate the need for
exotic matter in wormhole formation. We have shown that the
appropriate values of the parameters can maintain the weak energy
conditions without the need for exotic matter. This is a significant
finding as the existence of traversable wormholes would require the
presence of exotic matter with negative energy density and negative
pressure. The $f(Q,T)$ model suggests an alternative approach that
explores the effects of the accelerated expansion of the universe on
wormholes. By introducing viscosity, the model offers potential
explanations for the stability and formation of wormholes within the
framework of modified gravity. The numerical analysis conducted in
the paper indicates that the supporting matter of the wormhole is
near normal matter, which is a promising result. The analysis also
highlights the importance of energy conditions in studying the
properties of spacetime and the matter sources that generate
wormholes. In the following, weak energy conditions were
investigated for certain setups of the parameter space of the model.
It was shown that some of these fine-tunings establish weak energy
conditions near the wormhole. According to the selection of the
appropriate $\omega$ and also the determination of the  model
parameters $\alpha,\beta$, it is possible to establish the energy
conditions near the wormhole($\alpha <0 ~,~ 12.56 <\beta <25.12$ or
$\alpha >0 ~,~ \beta
>25.12 $), and some choices on the parameter space lead to the
violation of the weak energy conditions. \\
However, we offered a new perspective on wormhole solutions and
provided a model for further research in the field of modified
gravity and cosmology. The findings of this study have implications
for our understanding of the universe and the possibility of
traversable wormholes. The potential elimination of exotic matter in
wormhole formation opens up new ways for research and exploration in
the field of cosmology. Further research and exploration is needed
to unravel all components of wormholes and their implications in
cosmology.\\

{\bf  Declaration of interests}\\
The authors declare that they have no known competing financial
interests or personal equationships that could have appeared to
influence the work reported in this paper.

{\bf  Data availability}\\
 No new data were created or analyzed in this study.\\

\end{document}